\documentclass[twocolumn,showpacs,preprintnumbers,amsmath,amssymb]{revtex4}

\usepackage{graphicx}
\usepackage{dcolumn}
\usepackage{bm}

\begin{document}


\title{Evolution of Fishtail-effect with aging in pure and Ag-doped MG-YBCO}
\author{D. A. Lotnyk}
\email{dmitry.a.lotnik@univer.kharkov.ua}
\author{R. V. Vovk}
\author{M. A. Obolenskii}
\author{A. A. Zavgorodniy}
\affiliation{Physical department, V.N. Karazin Kharkov National University, 4 Svoboda Square, 61077 Kharkov, Ukraine.}
\author{J. Kov\'a\v{c}}
\author{M. Ka\v{n}uchov\'a}
\author{M. \v{S}efcikov\'a}
\author{V. Antal}
\author{P. Diko}
\affiliation{Material Physics Laboratory, Institute of Experimental Physics, Slovak Academy of Sciences, Watsonova 47, 04001 Ko\^{s}ice, Slovakia}
\author{A. Feher}
\affiliation{Centre of Low Temperature Physics, P.J. \^{S}afarik University, Park Angelinum 9, 041 54 Ko\^{s}ice, Slovakia}
\date{\today}
\pacs{74.25.Qt, 74.72.Bk, 74.81.Bd}

\begin{abstract}
$M(B)$-curves were experimentally investigated. Fishtail-effect (FE) was observed in MG YBa$_2$Cu$_3$O$_{7-\delta}$ and YBa$_2$Cu$_{3-x}$Ag$_x$O$_{7-\delta}$ (at $x \approx$ 0.02) crystals in a wide temperature range 40~K~$<T<$ 75~K at the orientation of magnetic field $\textbf{H}\parallel c$. It was obtained that the influence of bulk pinning on FE is more effective at low temperatures while the influence of surface barriers is more effective at high temperatures. The value $H_{max}$ for Ag-doped crystals is larger than for a pure one that due to the presence of additional pinning centers, above all on silver atoms.
\end{abstract}
\maketitle

\section{Introduction}
Since the discovery of high-temperature superconductors
(HTSCs), their engineering applications at liquid nitrogen
temperatures have drawn much attention.
The YBa$_2$Cu$_3$O$_{7-\delta}$ (YBCO) is one of the promising HTSCs materials
 for various technical applications. However, in ceramic materials the critical current density
($J_c$) is, due to the weak link effects, very low, rendering them unsuitable for applications. In
order to enhance $J_c$, the so-called melt-textured growth
(MG) process \cite{Jin88} has been developed, which can significantly
enhance $J_c$. One interesting phenomenon in the
MG HTSCs is the fishtail effect (FE) in $J_c$ as well as in
isothermal magnetic hysteresis loops ($M(H)$ curves) \cite{Muralidhar02, Koblischka00, Muralidhar03}.
As for the FE origin, some researchers propose the vortex
ordered phase to disordered phase transition \cite{Henderson96, Paltiel00, PaltielPRL00} to explain this interesting phenomenon, while others attribute it to the bulk pinning \cite{Muralidhar02, Koblischka00, Muralidhar03}. In the vortex phase transition
explanation, the FE exists both on a single $J_c(T)$ curve and
on a single $J_c(H)$ curve. However, in the bulk pinning explanation,
the FE appears on $M(H)$ curves whereas on a
single magnetization vs. temperature $M(T)$ curve it is seldom
reported.
In the previous work, the critical state model is used to
study the isothermal $M(H)$ curves \cite{Johansen97}. However, for
HTSCs, due to their operation at higher temperatures
and due to their small activation energy $U$, the flux creep is significant.
Hence, non-linear flux creep models were
developed. Meanwhile, the surface barrier \cite{Bean64}, which prevents
vortex entering ($H_{en}$) and exiting ($H_{ex}$) superconductors,
has strong effects on the irreversibility of
HTSCs. The goal of this work was to contribute to the
understanding of the FE origin in terms of significant influence of both bulk pinning and surface barriers.
\section{Experiment details and samples}
MG samples YBa$_2$Cu$_3$O$_{7-\delta}$ and YBa$_2$Cu$_{3-x}$Ag$_x$O$_{7-\delta}$ were grown as was described in \cite{Diko08}. Two samples, pure crystal (S1) and Ag-doped crystal (S2), were used in the present work. Physical properties of the samples are presented in  Table~\ref{tab:1}.
\begin{table}[h]
\begin{tabular}{|c|c|c|c|c|} \hline
Sample & $T_c$, K & $\Delta T_c$, K & a(b)$\times$b(a)$\times$c, mm$^3$ & m, mg\\ \hline
S1 & 90.2-89.8 & 1.0 & 2$\times$1.8$\times$0.7 & 12.2\\ \hline
S2 & 91.3-91.1 & 1.5 & 1.7$\times$1.6$\times$0.8 & 17.98\\ \hline
\end{tabular}
\caption{\label{tab:1}Physical characteristics for samples YBa$_2$Cu$_3$O$_{7-\delta}$ (S1) and YBa$_2$Cu$_{3-x}$Ag$_x$O$_{7-\delta}$ (S2)}
\end{table}
\begin{figure}
\includegraphics[clip=true,width=3.2in, height=3.5in]{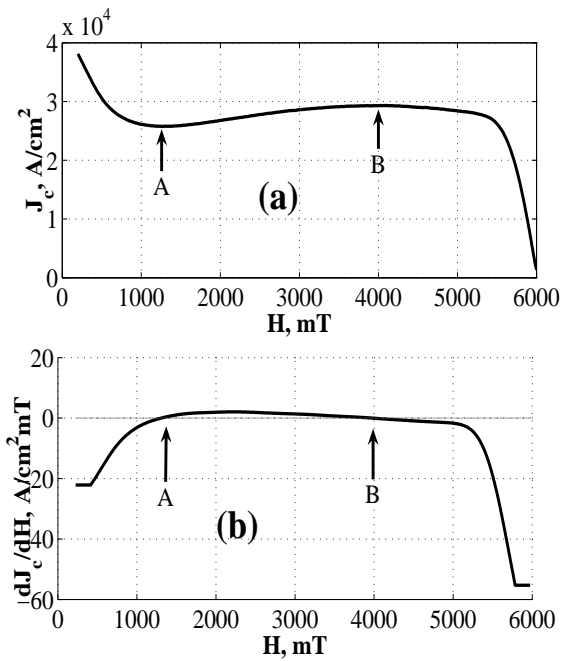}
\caption{\label{fig:1} Field dependences of critical current density $J_c$ (a) and derivative $dJ_c/dH$ (b) at temperature $T=$70~K for sample S1 right after annealing. Points A and B denotes positions of minimum and maximum respectively.}
\end{figure}
The measurements of isothermal hysteresis loops ($M(H)$ curves) at
various temperatures (20~K$<T<$85~K) were carried out in fields (-6~T$<H<$6~T) parallel to the c-axis by means of a commercial vibrating sample magnetometer (VSM). The field sweep rate was $dH/dt\approx 0.1$~T/min. To obtain an additional parameter $\Delta T_c$ the resistivity $R(T)$ curves were measured on two samples (the first one is pure and the second one is Ag-doped) which were obtained from the same single-domain bulks as S1 and S2. Transport measurements were performed using a physical property measurement system (PPMS). The experiment was divided into two steps.  $M(H)$ curves were measured: a) right after annealing of samples in oxygen (during 240 hours at 400$^{\circ}$C); b) after long-term aging at the room temperature (six months), in which case almost all relaxation processes were finished. The first value for $T_c$ in Table~\ref{tab:1} corresponds to a) and the second value corresponds to b).
\section{Model to analyze experimental data}
From the measured $M(H)$ curves the critical current density was calculated using the critical state model \cite{Bean62} that have been developed in \cite{Wiesinger92}:
\begin{eqnarray}
{J_c = 2\Delta M\frac{\rho}{a^2\left(b-a/3\right)c}},
\label{eq:1}
\end{eqnarray}

where $\Delta M=M_1(H)-M_2(H)$ ($M_1(H)>$0, $M_2(H)<$0 are the values of magnetic moment at field $H$, $\rho$ is the density of sample, a,b and c are geometric dimensions of sample. Dependences $J_c(H)$ (Fig.~\ref{fig:1}a for $T$=70~K) were calculated for the field range 0~T$<H<$6~T. To obtain the exact position of maximum on $J_c(H)$ curves the derivative $dJ_c/dH$ was calculated for each temperature (see example at Fig.~\ref{fig:1}b). Fig.~\ref{fig:1} shows positions of both minimum (point A) and maximum (point B) on J vs H curve, where derivative $dJ_c/dH=$0.
\begin{figure}
\includegraphics[clip=true,width=3.2in]{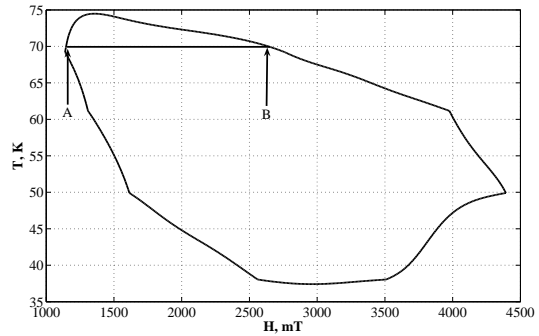}
\caption{\label{fig:2} Contour for $dJ_c/dH=0$ for sample S1 right after annealing. Points A and B coincide with points A and B in the Fig.~\ref{fig:1}}
\end{figure}
Then the 3D surface $T-H-dJ_c/dH$ was obtained. Intermediate curves were defined by the 3rd degree polynomial (spline-function). The contour for $dJ_c/dH=0$ is shown in Fig.~\ref{fig:2}. Points A and B correspond to those in Fig.~\ref{fig:1}. This method is useful to obtain position of maximum $J_{c,max}$ (or ensemble of points B) not only at given temperatures but also for intermediate temperatures.
\section{Results and discussions}
\begin{figure}
\includegraphics[clip=true,width=3.2in]{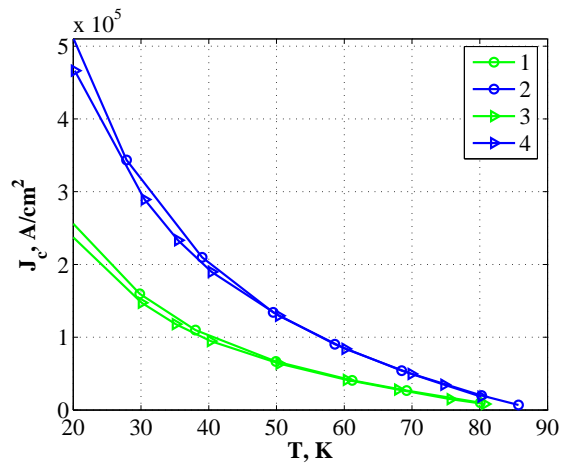}
\caption{\label{fig:3} Temperature dependences of critical current density at zero magnetic field for samples S1 (green lines) and S2 (blue lines), right after annealing (curves 1,2) and after long-term aging (curves 3,4)}
\end{figure}
\begin{figure}
\includegraphics[clip=true,width=3.2in]{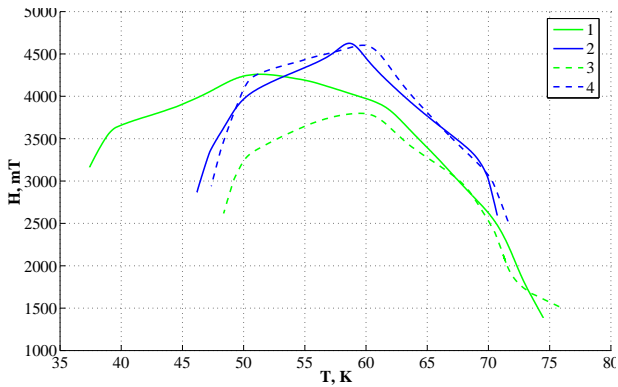}
\caption{\label{fig:4} Temperature-field dependences of $J_{c,max}$-position for samples S1 (green lines) and S2 (blue lines), right after annealing (curves 1,2) and after long-term aging (curves 3,4)}
\end{figure}

Fig.~\ref{fig:3} shows the temperature dependences $J_c(T)$, on which FE appears. However, according to \cite{Henderson96, Paltiel00, PaltielPRL00}, FE appears not due to the order-disorder phase transition of vortex lattice. The possible reason of the FE existence is a significant influence of both bulk pinning and surface barriers \cite{Zhang06}. From the model in previous chapter the $H-T$ dependences of $J_{c,max}$ were obtained (see Fig.~\ref{fig:4}).
As one can see from Table~\ref{tab:1} the value of the critical temperature $T_c$ for sample S1 changed from 90.2~K to 89.8~K and for sample S2 changed from 91.3~K to 91.1~K. This indicates the optimal oxygen annealing process and aging effects are not caused by changing in $\delta$ (especially for sample S2). According to the estimations performed in \cite{Zhang06} the surface barriers are more important at high temperatures while the bulk pining is more important at low temperatures. As one can see from Fig.~\ref{fig:4} the position of $J_{c,max}$ significantly evolves after long-term aging. The maximum field $H_{max}$ for sample S1 decreases from 4.25~T to 3.8~T and the minimum temperature $T_{min}$ increases from 37.5~K to 48~K (curves 1 and 3 in Fig.~\ref{fig:4}). Such changes can be induced by the redistribution of pinning centers in the crystal or, in another words, bulk pinning changes during aging. Constancy of the maximum temperature $T_{max}$ for S1 corresponds to the influence of surface barriers at high temperatures. The aging effect for sample S2 is not so significant as for S1. It is related to the presence of considerable mechanical tensions in the sample (large value of an additional parameter $\Delta T_c=$1.5~K) that prevent the redistribution of pinning centers. Larger value of $H_{max}$ (for S2 as compared with S1) corresponds to the presence of additional pinning centers on silver atoms \cite{Nakashima08}.

A possible reason of the decrease of $H_{max}$ and increase of $T_{min}$ for S1 is the decreasing number of dislocations. In another words, the crystal S1 becomes more homogeneous after aging.
\section{Conclusions}
In summary, MG YBCO samples were investigated by measuring $M(B)$-curves. On these curves FE was observed is in a wide temperature range. The origin of the FE existence is a significant influence of both surface barrier and bulk pinning. Bulk pinning influences parameters $H_{max}$ and $T_{min}$ while surface barrier influences $T_{max}$. Undoped YBCO is characterized by a significant aging effect which is manifested in a considerable change of $H_{max}$ and $T_{min}$. This corresponds to the change of bulk pinning. For Ag-doped YBCO there are no significant changes in $H_{max}$, $T_{min}$ and $T_{max}$. Such difference is caused by the redistribution of pinning centers in undoped YBCO while Ag-doped YBCO is characterized by huge mechanical tensions which prevent redistribution. Possible origin of redistribution for S1 is a decrease in number of dislocations, i.e. sample becomes more homogenous.

\textit{Acknowledgments}

The work was partly supported by the Slovak Research and Development  Agency (No. APVV-0006-07) and  the Slovak Grant Agency VEGA (No.1/0159/09) The financial support of U.S.Steel - DZ Energetika Ko\v{s}ice is acknowledged. One of author (D.L.) is very thankful to the National Scholarship Program of Slovak Republic (SAIA) for the financial support during his stay in Slovakia.


\end{document}